\documentclass{ws-procs10x7}

\newcommand{\nn}{\nonumber} 

\newcommand{\nslash}{n\!\!\!\!\slash}

\begin{document}

\title{Factorization in color-suppressed $\bar B\to D^{(*)}\pi$ decays 
from the Soft-Collinear Effective Theory}

\author{Dan Pirjol}

\address{Center for Theoretical Physics, 77 Massachusetts Ave., Cambridge, 
MA 02139\\E-mail: pirjol@lns.mit.edu}

\twocolumn[\maketitle\abstract{
The soft-collinear effective theory has been recently applied to prove
novel factorization theorems for many $B$ decay processes. We describe here
in some detail the factorization relation for color-supressed nonleptonic 
$B\to D^{(*)0}\pi^0$ decays and update the phenomenological analysis of 
these decays.}]

\section{Factorization and SCET}

The application of factorization to exclusive processes has been around
now for 25 years (see \cite{BLreview} for a review of the early literature). 
These factorization theorems have been traditionally 
proved using diagrammatic methods in 
perturbation theory. Apart from being rather involved technically, this
 approach has met with very limited success beyond leading order
in the $1/Q$ expansion, where power suppressed
terms generally give rise to divergent convolution integrals.

Recently an alternative approach to factorization in exclusive
processes has been proposed, based on the soft-collinear effective theory (SCET)
\cite{Bauer:2000ew}. This greatly simplifies the proof of factorization
theorems and allows a systematic treatment of power corrections.

The SCET is constructed as a systematic expansion in 
$\lambda = \Lambda/Q\ll 1$, where $Q$ is a large scale specific
to the problem, such as $Q^2 = -q^2$  in the electromagnetic $\pi$ form factor 
at a space-like momentum transfer, or $Q \sim m_b$ in the heavy quark decay,
and $\Lambda \sim 500$ MeV is the QCD scale. This is achieved by identifying
the relevant energy scales and the corresponding degrees of freedom.
The Lagrangian and operators of SCET (such as currents) are
organized in a series in $\lambda$ as ${\cal L } = {\cal L}^{(0)} + {\cal L}^{(1)}+
\cdots$, etc.

The exclusive processes considered here receive contributions from  
3 well-separated scales $Q^2 \gg
\Lambda Q \gg \Lambda^2$. This requires the introduction of a sequence of
effective theories QCD $\to$ SCET$_{\rm I} \to$ SCET$_{\rm II}$, which contain
degrees of freedom of successively lower virtuality.
The intermediate theory ${\rm SCET}_{\rm I}$ contains hard-collinear quarks
and gluons with virtuality $p_{\rm hc}^2 \sim \Lambda Q$ and ultrasoft
partons with virtuality $\Lambda^2$.
The final theory ${\rm SCET}_{\rm II}$ includes only soft and collinear modes
with virtuality $p^2 \sim \Lambda^2$.

Using the SCET many new factorization theorems were derived (see e.g.~\cite{IWS}
for other recent applications). 
I will describe in 
this talk a factorization theorem for exclusive color-suppressed $B^0\to D^0\pi^0$
decays~\cite{Mantry:2003uz}.

\begin{figure}
\includegraphics[height=.09\textheight]{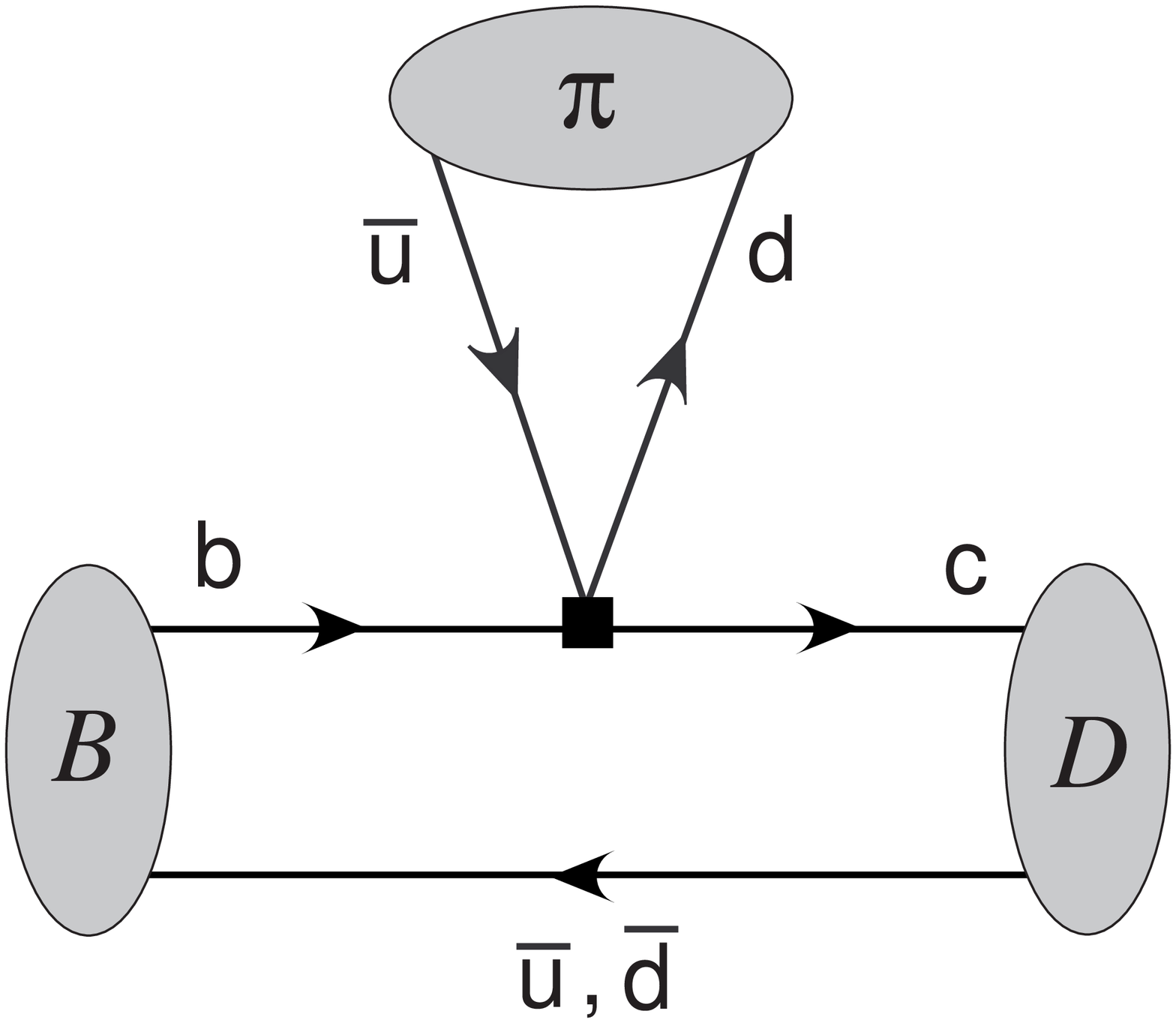}
  \includegraphics[height=.09\textheight]{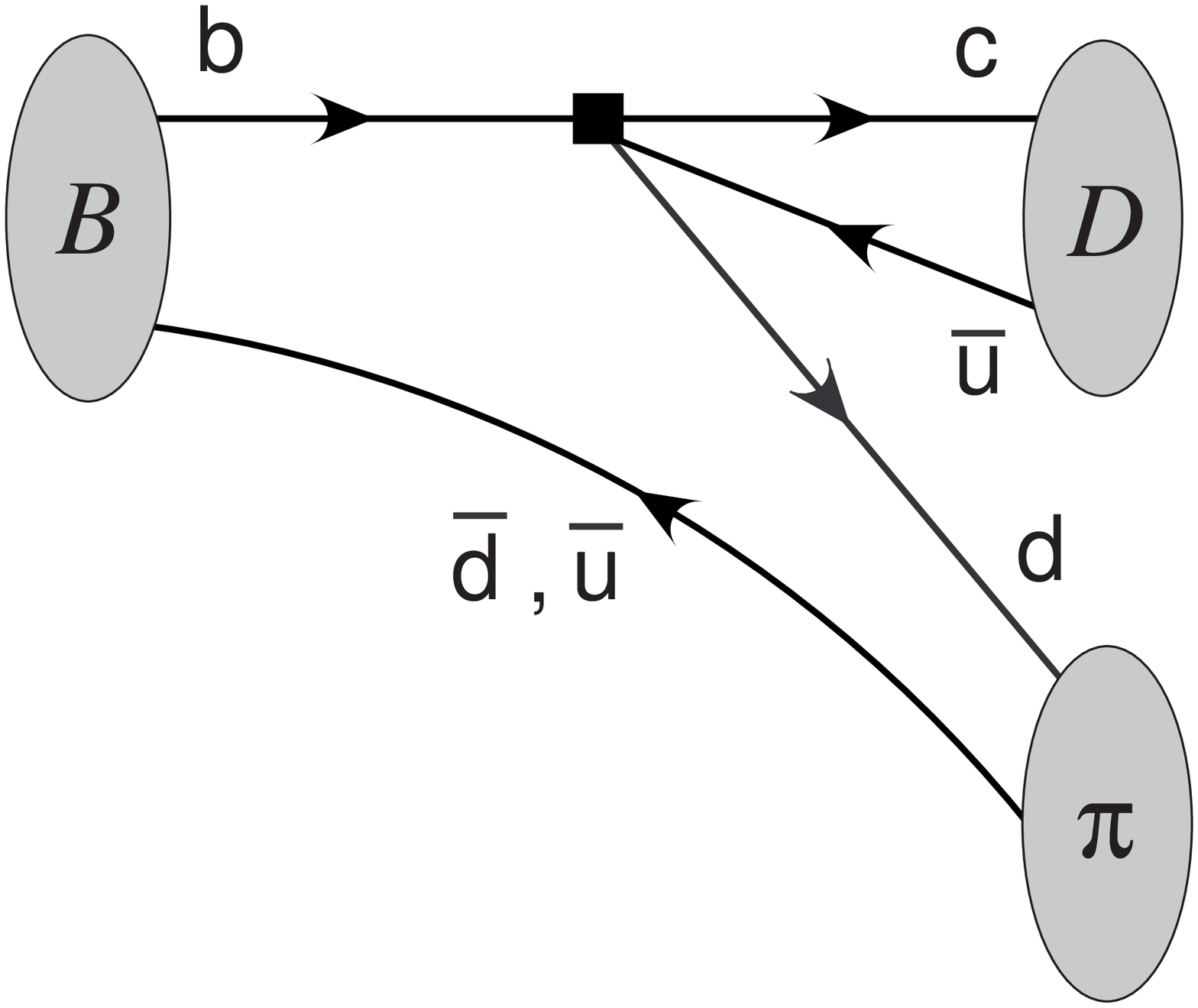}
  \includegraphics[height=.09\textheight]{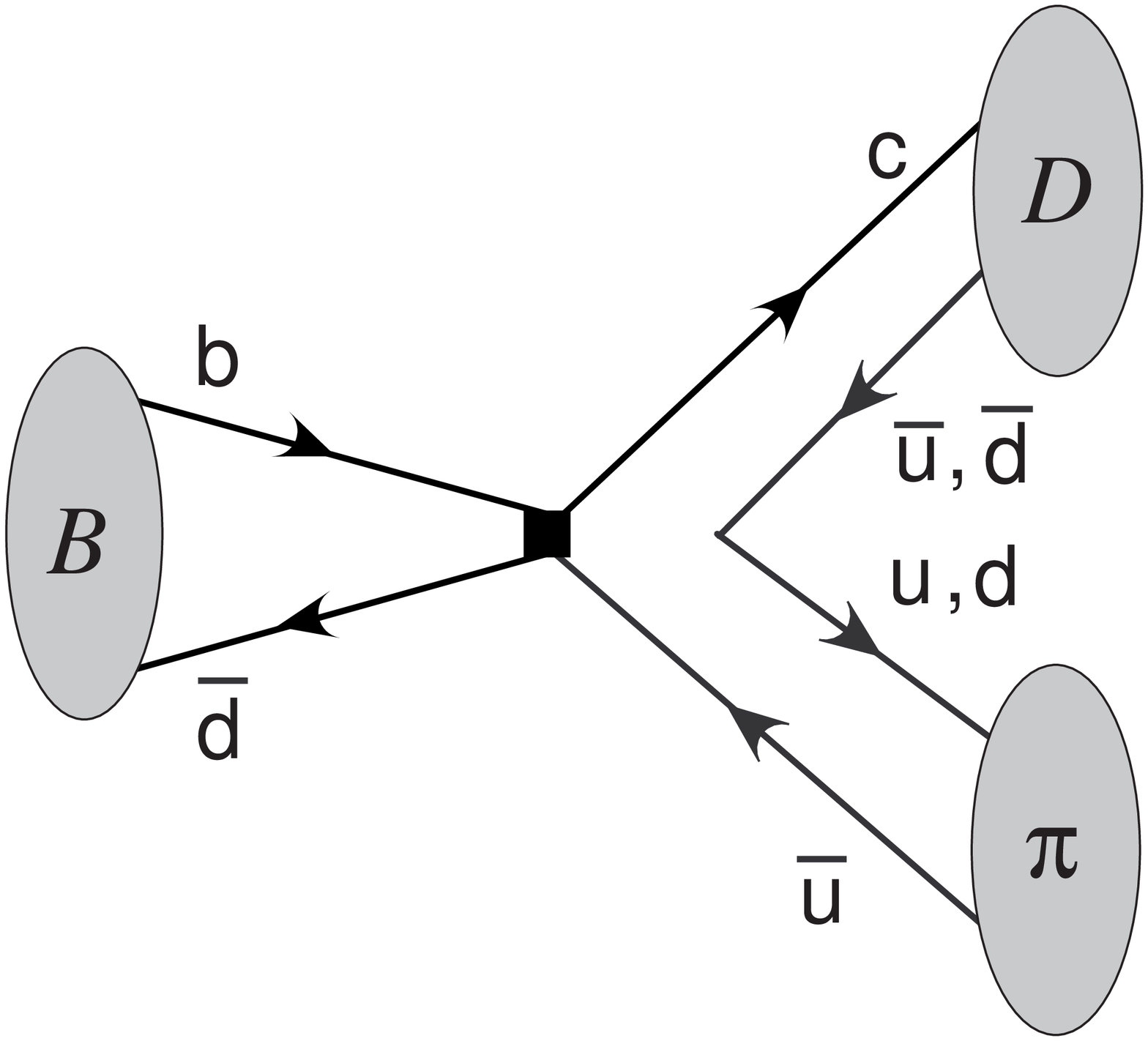}
  \caption{$\bar B\to D\pi$ topologies: T, C, and E. 
Only C and E contribute to color suppressed decays.}
  \label{fig:cs}
\end{figure}

\section{Factorization in color-suppressed $B\to D\pi$ decays}

The $\bar B\to D\pi$ decays  are mediated by the weak Hamiltonian
\begin{eqnarray*} \label{Hw}
  H_w = \frac{G_F}{\sqrt2} V_{cb} V_{ud}^* \big[ C_1 (\bar c b)
      (\bar d u) + 
    C_2 (\bar c_i b_j) (\bar d_j u_i) \big] \,,
\end{eqnarray*} 
which can contribute through the flavor contractions shown in 
Fig.~\ref{fig:cs}.
Denoting the amplitudes as $A_{+-}=A(\bar B^0\to D^+\pi^-)$, $A_{0-}=A(B^-\to
D^0\pi^-)$, and $A_{00} = A(\bar B^0 \to D^0\pi^0)$, one has the decomposition
into graphical and isospin amplitudes~\cite{Rosner}
\begin{eqnarray}
  A_{+-} &=& T+E = \frac{1}{\sqrt{3}} A_{3/2} + \sqrt{\frac23} A_{1/2} \,,\nn\\
  A_{0-} &=& T+C = \sqrt{3}\, A_{3/2} \,, \\
  A_{00} &=& \frac{C-E}{\sqrt{2}} 
     =\sqrt{ \frac23} A_{3/2} - \frac{1}{\sqrt3} A_{1/2} 
     \,. \nn
\end{eqnarray}

\begin{table*}[t!]
\begin{center}
\begin{tabular}{|c|c|c|c|c|c|}
\hline
 Decay & Br$(10^{-3})$ & $(|R_I|, \delta_I)$
 & Decay & Br$(10^{-3})$ & $(|R_I|, \delta_I)$
 \\
\hline\hline
$\bar B^0 \to D^+ \pi^-$ & $2.76\pm 0.25$ & $0.70 \pm 0.07$ 
 & $\bar B^0 \to D^{*+} \pi^-$ & $ 2.76 \pm 0.21 $  & $0.76 \pm 0.07$
 \\ 
$B^- \to D^0 \pi^-$ &  $4.98\pm 0.29$  & $28.1^\circ\pm 3.3^\circ$
 & $B^- \to D^{*0} \pi^-$ & $ 4.6 \pm 0.4$   & $31.9^\circ \pm 4.5^\circ$
 \\ 
$\bar B^0 \to D^0 \pi^0$ & $0.260\pm 0.022$ &  
& $\bar B^0 \to D^{*0} \pi^0$   & $0.27 \pm 0.05$ & 
  \\
\hline\hline
\end{tabular}
\caption{Experimental data on $\bar B\to D^{(*)}\pi$ and the corresponding
results for the ratio of isospin amplitudes $R_I = A_{1/2}/(\sqrt2 A_{3/2}) \equiv
|R_I| e^{i\delta_I}$. The data is taken from [12], except for $D^0\pi^0$ which 
is
the average of [11]. We use $\tau(B^+)/\tau(B^0) = 1.086\pm 0.017$ [12].}
\end{center}
\end{table*}

The color-allowed amplitudes $A_{+-}$ and $A_{0-}$ are
described by a factorization
theorem \cite{Dugan:1991de,Politzer:1991au,Beneke:2000ry}, proven with
SCET in \cite{Bauer:2001cu}
\begin{eqnarray*} \label{Alo}
  A^{(*)} = N^{(*)}\: \xi(w_0) \int_0^1\!\!dx\: 
   T^{(*)}(x,\mu)\: \phi_\pi(x) +\ldots ,
\end{eqnarray*}
where $\xi(w_0)$ is the Isgur-Wise function at maximum recoil, $\phi_\pi(x)$
is the light-cone distribution function for the pion, $T=1+ O(\alpha_s)$ is the
hard scattering kernel, and $N^{(*)}= \frac{G_F}{\sqrt{2}} V_{cb} V_{ud}^* E_\pi
f_\pi \sqrt{m_{D^{(*)}} m_B}(1+m_B/m_{D^{(*)}})$. The ellipses in
Eq.~(\ref{Alo}) denote terms suppressed by $\Lambda/Q$ where
$Q=\{m_b,m_c,E_\pi\}$. The predictions from Eq.~(\ref{Alo})
are in good experimental agreement with data on color allowed decays.

Large $N_c$ QCD gives an alternative justification for factorization in
color allowed decays. Possible tests for the underlying factorization mechanism
use decays into multibody states $B\to D^{(*)} n\pi$~\cite{Ligeti:2001dk},
isospin analyses of such decays ~\cite{Bauer:2002sh,Zito} and decays into 
hadrons with exotic quantum numbers~\cite{Diehl:2001xe}. 

The color-suppressed amplitude $A_{00}$ gets contributions from $C$ and $E$, but
not $T$.  Large $N_c$ predicts $C/T\sim E/T\sim 1/N_c$ (counting $C_1\sim 1$ and
$C_2\sim 1/N_c$). Writing the isospin relation among amplitudes as
\begin{eqnarray} \label{rel}
  R_I  = 1 - \frac{3}{\sqrt{2}} \frac{A_{00}}{A_{0-}} \,,
\end{eqnarray}
where $R_I = A_{1/2}/(\sqrt{2}A_{3/2}) \equiv |R_I| e^{i\delta_I}$, 
the suppression of $C,E$ is measured by the deviation of $R_I$ from 1 in
the complex plane~\cite{Neubert:2001sj}. We show in Table I the present
experimental situation for $\bar B\to D^{(*)}\pi$ decays, together with the
corresponding results for $R_I$. 

\begin{figure}
  \includegraphics[height=.09\textheight]{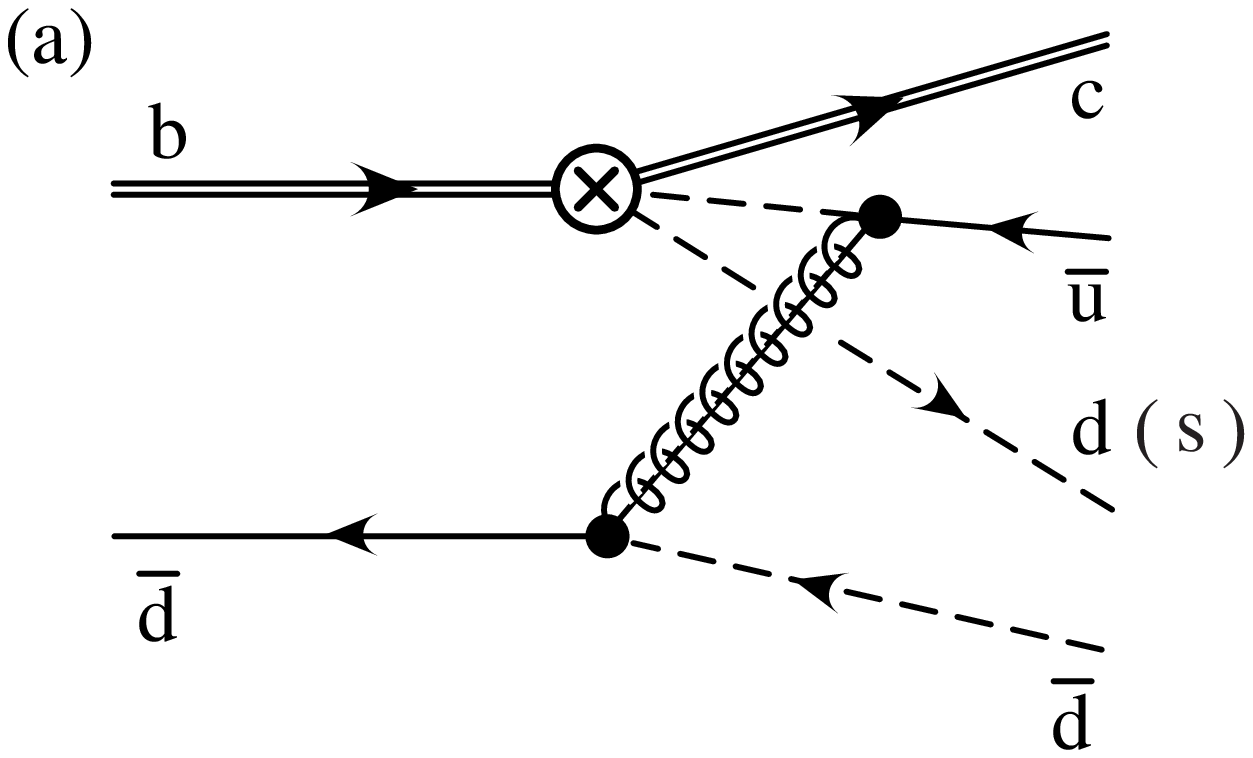}
  \includegraphics[height=.09\textheight]{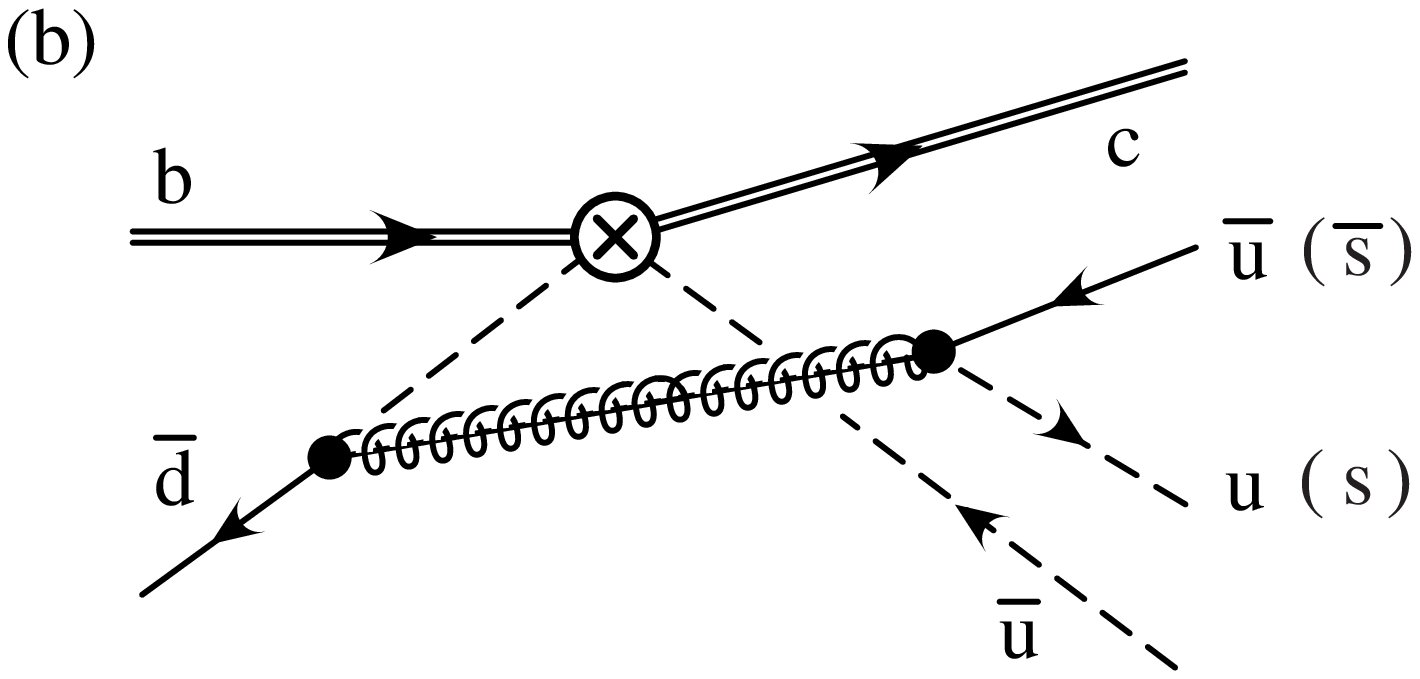}
    \caption{Diagrams in ${\rm SCET}_{\rm I}$ for tree level matching. 
      $\otimes$ denotes the operator $Q_j^{(0,8)}$ and the dots are insertions
      of ${\cal L}_{\xi q}^{(1)}$. The solid lines and double solid lines carry
      momenta $p^\mu\sim \Lambda$ and form the $B$ and $D$. The dashed lines are
      energetic collinear quarks that form the light meson $M$.}
  \label{fig:I}
\end{figure}

SCET gives a quantitative description of  the color suppressed amplitude, expressed
as a factorization theorem for $A_{00}$. The main observation is that in SCET$_{\rm I}$
these decays are mediated by only one type of operators~\cite{Mantry:2003uz}
${\cal H}_W \to T( Q_{j}^{(0,8)}(0), i {\cal L}_{\xi q}^{(1)}(x), i
{\cal L}_{\xi q}^{(1)}(y))$. By power counting it follows that these decays are 
power suppressed
by $\Lambda/Q$, with $Q = \{ m_c, m_b, E_\pi \}$.
This T-product contributes
at tree level as shown in Fig.~\ref{fig:I}.

When matched onto ${\rm SCET}_{\rm II}$ the time-ordered product gives a product
of soft $O_s^{(0,8)}$ and collinear $O_c$ operators, where the soft operators are 
\begin{eqnarray}\nonumber
  O_s^{(0)} \!\!\! &=& \!\!\!
  (\bar h_{v'}^{(c)} S)  \nslash P_L \, (S^\dagger h_v^{(b)}) \:
  (\bar d\,S)_{k^+_1} \nslash P_L\,  (S^\dagger u)_{k^+_2} 
  \,,
\end{eqnarray}
and $O_s^{(8)}$ is identical with color structure $T^a\otimes T^a$. 
In addition there are operators encoding long-distance contributions in ${\rm
  SCET}_{\rm II}$ coming from the
region of momentum space in Fig.~\ref{fig:I} with a hard gluon ($p_g^2\sim
Q\Lambda$), but an on-shell quark propagator ($p_q^2\sim \Lambda^2$).

Heavy quark symmetry relates the $\bar B\to D$ and $\bar B\to D^*$ matrix elements of 
$O_s^{(0,8)}$ as
\begin{eqnarray} \label{S}
  \langle D^{(*)} | O_s^{(0,8)} | B \rangle = \sqrt{m_B m_D^{(*)}}
S_L^{(0,8)}(k_1^+,k_2^+) \,,
\end{eqnarray}
The complete factorization theorem for color suppressed decays can now be written as
~\cite{Mantry:2003uz}
\begin{eqnarray}\label{fact}
&& A_{00}^{D^{(*)}M} \!\! = \!\!
   N_0^{M} \int\!\! dx\, dz\, d{k_i^+}\, 
   T_{L\mp R}^{(i)}(z)\\
&& \times J^{(i)}(z,x,k_i^+)
 S^{(i)}(k_1^+,k_2^+)\: \phi_M(x) + A_{long}^{D^{(*)}M}\,.\nn
\end{eqnarray}
with $T_{L\mp R}^{(i)}$ hard scattering kernels and $N_0^{M}= \frac{G_F}{2}
V_{cb}V_{ud}^* f_M \sqrt{m_B m_{D^{(*)}}}$. The non-perturbative dynamics is
encoded in $\phi_M$, the light-cone distribution function of the light meson $M$, and
two soft functions for the $B\to D^{(*)}$ transition $S^{(0,8)}(k_i^+)$ with 
$k_i^+$ the light spectator momenta. The jet function $J^{(i)}$ 
appears in the matching SCET$_{\rm I} \to $ SCET$_{\rm II}$ and contains the 
effects from scales $p_{\rm hc}^2 \sim E_\pi\Lambda$. There is a further simplification
for $M$ an isovector meson such as $\pi,\rho$, for which the amplitude 
$A_{long}$ vanishes from $G$ invariance.

We discuss next several important implications of this factorization theorem. 
The soft functions $S_L^{(0,8)}$ are complex, encoding 
final state interactions arising from the soft Wilson line along $n$ 
in the definition of the soft operators. This is a source of nonperturbative
final state interaction effects, similar to those producing single
spin asymmetries~\cite{SSA} in semi-inclusive DIS.

Second, since heavy quark symmetry relates the soft functions in $B\to D^{(*)}$, 
one predicts a relation between these decays
\begin{eqnarray} \label{p1}
&&  \delta_I(D^*\pi) = \delta_I(D\pi) + O(\alpha_s(Q),\frac{\Lambda}{Q})\,,\\
&&  A_{00}^{D^{*}\pi} =
A_{00}^{D\pi} + O(\alpha_s(Q),\frac{\Lambda}{Q}) 
      \,.  \nn
\end{eqnarray}
We emphasize that heavy quark symmetry alone would not have been sufficient to
make this prediction, which requires soft-collinear factorization as a 
crucial ingredient. The current experimental
data (see Table I) is in good agreement with Eq.~(\ref{p1}).  

Additional predictions are possible by expanding the jet functions in 
perturbation theory and working at leading order in $\alpha_s(\mu_C)$ with
$\mu_C^2 = \Lambda Q$.
The convolution integral over $x$ can now be performed exactly which gives
\begin{eqnarray} \label{p4}
&&  A_{00}^{D^{(*)}M} \!\! =\!\!
   \frac{G_F}{2}V_{cb} V^*_{ud} f_\pi \sqrt{m_B m_D}\\
&&\times (C_1+\frac{C_2}{N_c}) \frac{4\pi C_F\alpha_s(\mu_C)}{N_c} 
    s_{\rm eff} \langle x^{-1} \rangle_M \,,\nn
\end{eqnarray}
with $\langle x^{-1} \rangle_M = \int dx\:
\frac{\phi_M(x)}{x}$, and $s_{\rm eff}=-s^{(0)} + C_2/[N_c C_F (C_1+C_2/N_c)]\, s^{(8)}$ with
$s^{(0,8)} (\mu)= -\int dk_1^+ dk_2^+ \frac{1}{k_1^+ k_2^+}\: S^{(0,8)}(k_i^+,\mu)$.
Corrections to Eq.~(\ref{p4}) are $O(\alpha_s(\mu_C),\Lambda/Q)\simeq 30\%$. 
The strong phase Arg$(A_{00}/A_{-0})$ comes from $s_{\rm eff}$ 
and is thus independent on the light meson. This predicts the universality 
of the strong phase $\phi\equiv$ Arg$(A_{00}/A_{0-})$ for $D^{(*)}\pi$ and $D^{(*)}\rho$.
The data in the Table I gives
$s_{\rm eff} \equiv 
|s_{\rm eff}| e^{i\phi}$ with $(|s_{\rm eff}|,\phi) = (429\pm 18 \mbox{ MeV},
\pm(38^\circ\pm 12^\circ))$ at $\mu_c=2.31$ MeV where $\alpha_s(\mu_C) = 0.25, C_1 = 1.15,
C_2=-0.362$. We used here $|V_{cb}|=41.9\times 10^{-3}$.

Finally, the relation (\ref{p4}) gives the leading deviation from 1 of the ratios of
decays into charged pions
\begin{eqnarray} \label{p5}
&&   R_c^{DM} \equiv
  \frac{A(\bar B^0\to D^{+}M^-)}{A(B^-\to D^{0}M^-)} \\
&& = 1 - \frac{16\pi\alpha_s(\mu_C) m_D} {9 (m_B + m_D ) E_M}
     \frac{\langle x^{-1} \rangle_M}{\xi(\omega_0)}
s_{\rm eff}\,. \nn
\end{eqnarray}
Predictions for color-suppressed decays using other methods have been discussed
in Refs.~\cite{Chiang:2002tv}.

\section{Conclusions}

This talk described the application of SCET to derive a new factorization relation
for color suppressed decays $\bar B\to D^{(*)}\pi$. The result is different from 
the usual naive factorization {\em Ansatz}  for
color suppressed amplitudes \cite{BSW} and depends on a new
nonperturbative function describing the $B\to D$ transition. 

Similar factorization relations were derived for other $B$ decays into charm: 
nonleptonic baryon decays $\Lambda_b \to 
\Sigma_c \pi$~\cite{Leibovich:2003tw}, decays into orbitally
excited states $\bar B^0\to D^{**0}\pi^0$~\cite{Mantry:2004pg}, 
and decays into isosinglet states with an 
$\eta^{(\prime)},\omega,\phi$~\cite{Blechman:2004vc}.

Nonleptonic B decays to heavy-light states with charm have a rich phenomenology,
and the factorization theorem (\ref{fact}) can be expected to be an useful 
tool to organize and explore the implications of this data for
the nonperturbative structure of the heavy
quark systems.

\section*{Acknowledgments}

This work was supported by the
U.S. Department of Energy (DOE) under the Grant No.
DOE-FG03-97ER40546 and by the NSF under grant PHY-9970781.


\begin{thebibliography}{99}

\bibitem{BLreview} S.~J.~Brodsky and G.~P.~Lepage, 
Adv.\ Ser.\ Direct.\ High Energy Phys.\ {\bf 5}, 93 (1989).

\bibitem{Bauer:2000ew}
C.~W.~Bauer, S.~Fleming and M.~E.~Luke,
Phys.\ Rev.\ D {\bf 63}, 014006 (2001);
C.~W.~Bauer, S.~Fleming, D.~Pirjol and I.~W.~Stewart,
Phys.\ Rev.\ D {\bf 63}, 114020 (2001);
C.~W.~Bauer and I.~W.~Stewart,
Phys.\ Lett.\ B {\bf 516}, 134 (2001);
C.~W.~Bauer, D.~Pirjol and I.~W.~Stewart,
Phys.\ Rev.\ D {\bf 65}, 054022 (2002).

\bibitem{IWS} See the talks by I.~W.~Stewart and C.~W.~Bauer 
at this conference.

\bibitem{Rosner} J.~L.~Rosner, 
Phys.\ Rev.\ D {\bf 60}, 074029 (1999).

\bibitem{Mantry:2003uz}
S.~Mantry, D.~Pirjol and I.~W.~Stewart,
Phys.\ Rev.\ D {\bf 68}, 114009 (2003)


\bibitem{Dugan:1991de}
M.~J.~Dugan and B.~Grinstein,
Phys.\ Lett.\ B {\bf 255}, 583 (1991).


\bibitem{Politzer:1991au}
H.~D.~Politzer and M.~B.~Wise,
Phys.\ Lett.\ B {\bf 257}, 399 (1991).

\bibitem{Beneke:2000ry}
M.~Beneke, G.~Buchalla, M.~Neubert and C.~T.~Sachrajda,
Nucl.\ Phys.\ B {\bf 591}, 313 (2000).


\bibitem{Bauer:2001cu}
C.~W.~Bauer, D.~Pirjol and I.~W.~Stewart,
Phys.\ Rev.\ Lett.\  {\bf 87}, 201806 (2001).

\bibitem{Neubert:2001sj}
M.~Neubert and A.~A.~Petrov,
Phys.\ Lett.\ B {\bf 519}, 50 (2001).


\bibitem{D0pi0}
B.~Aubert {\it et al.} [BABAR Collaboration],
Phys.\ Rev.\ D {\bf 69}, 032004 (2004);
K.~Abe {\it et al.} [BELLE Collaboration],
arXiv:hep-ex/0409004.

\bibitem{PDG}
S.~Eidelman {\it et al.} [Particle Data Group Collaboration],
Phys.\ Lett.\ B {\bf 592}, 1 (2004).

\bibitem{Ligeti:2001dk}
Z.~Ligeti, M.~E.~Luke and M.~B.~Wise,
Phys.\ Lett.\ B {\bf 507}, 142 (2001).


\bibitem{Diehl:2001xe}
M.~Diehl and G.~Hiller,
JHEP {\bf 0106}, 067 (2001).


\bibitem{Bauer:2002sh}
C.~W.~Bauer, B.~Grinstein, D.~Pirjol and I.~W.~Stewart,
Phys.\ Rev.\ D {\bf 67}, 014010 (2003).

\bibitem{Zito}
M.~Zito,
Phys.\ Lett.\ B {\bf 586}, 314 (2004).

\bibitem{Coan:2001ei}
Coan, T.~E., et~al., \emph{Phys. Rev. Lett.}, \textbf{88}, 062001 (2002).

\bibitem{Abe:2001zi}
Abe, K., et~al., \emph{Phys. Rev. Lett.}, \textbf{88}, 052002 (2002).

\bibitem{Aubert:2003sw}
Aubert, B., et~al., \emph{arXiv:hep-ex/0310028} (2003).

\bibitem{Chiang:2002tv}
C.~W.~Chiang and J.~L.~Rosner,
Phys.\ Rev.\ D {\bf 67}, 074013 (2003);
Y.~Y.~Keum {\em et al}, 
Phys.\ Rev.\ D {\bf 69}, 094018 (2004);
C.~K.~Chua, W.~S.~Hou and K.~C.~Yang,
Phys.\ Rev.\ D {\bf 65}, 096007 (2002);
L.~Wolfenstein,
Phys.\ Rev.\ D {\bf 69}, 016006 (2004).

\bibitem{SSA} 
S.~J.~Brodsky, D.~S.~Hwang and I.~Schmidt, Phys.\ Lett.\ B {\bf 530}, 99 (2002).

\bibitem{BSW}
M.~Bauer, B.~Stech and M.~Wirbel, Z.\ Phys.\ C {\bf 34}, 103 (1987).

\bibitem{Leibovich:2003tw}
A.~K.~Leibovich, Z.~Ligeti, I.~W.~Stewart and M.~B.~Wise,
Phys.\ Lett.\ B {\bf 586}, 337 (2004)

\bibitem{Mantry:2004pg}
S.~Mantry,
arXiv:hep-ph/0405290.

\bibitem{Blechman:2004vc}
A.~E.~Blechman, S.~Mantry and I.~W.~Stewart,
arXiv:hep-ph/0410312.

\end{thebibliography}
\end{document}